# Stochastic mutation as a mechanism for the emergence of SARS-CoV-2 new variants -- A Scientific Conjecture on Artificial Intelligence Paradigm

Liaofu Luo [a], Jun Lv [b]

[a)] Faculty of Physical Science and Technology, Inner Mongolia University, Hohhot 010021, China

[b)] College of Science, Inner Mongolia University of Technology, Hohhot 010051, China

## Abstract

The article summaries authors' researches on spreading dynamics of COVID-19 by use of the method of continuously asking and answering questions. Scientific conjecture provides the logical starting point and the important driving force of scientific research. The present article also proposes a scientific conjecture on the development of artificial intelligence: Question-answering AI is highly efficient and low-cost as compared with AI agent.

**Keywords:** COVID-19, spreading dynamics, scientific conjecture, question-answering AI

By using the method of Response-to-continuous-questions we shall summary our studies in the following on the spreading dynamics of COVID-19 with the central problem on the stochastic mutation as a mechanism for the emergence of SARS-CoV-2 new variants.

Question-answering AI（for example ChatGPT）and self-service AI (AI agent) represent two distinct paradigms of artificial intelligence. Their fundamental distinction lies in whether the system operates through passive guidance or proactive execution. The deployment cost of AI Agent is significantly greater than that of question--answering AI. The key to scientific research lies in asking the right questions. If one can continuously ask clear, well-defined questions step by step, question-answering AI is highly efficient and low-cost. The following summary may provide an example of question-answering AI.

## 1, How to formulate a mathematical representation of a SARS-CoV-2 mutant [2][3]

For each mutation at residue i, the mean and standard deviation of the binding affinity are denoted as $b_i$ and $s_i$, respectively. A mutation is classified as affinity-enhancing if its measured affinity $m_i$ satisfies $m_i>b_i+s_i$; affinity-weakening if $m_i<b_i- s_i$; and negligible in effect if $b_i- s_i\leq m_i\leq b_i+s_i$. The surface of coronaviruses is decorated with a spike protein, which in SARS-CoV-2 consists of approximately 1273 amino acids. Each SARSCoV-2 strain is represented by a four-letter sequence, where each letter indicates the mutation type at a specific residue. The encoding is defined as follows: 0 denotes no mutation relative to the wild type; 1 indicates a mutation with negligible effect on ACE2 binding affinity; 2 represents an affinity-enhancing mutation; and 3 signifies an affinity-weakening mutation. Thus, the full spike protein of any SARS-CoV-2 mutant is represented by a 1273-character sequence using this four-letter alphabet. Note that for regions outside the RBD, only the symbols 0 and 1 are used.

The four-letter scheme uses binary values (0 and 1) for 1273 mutational sites on the spike protein, with two additional options (2 and 3) permitted at the 201 amino acid residues on the receptor-binding domain (RBD). Due to interactions between the RBD and the human ACE-2 receptor, the 201 amino acid residues in the RBD cannot vary independently (indicated by Sultana et al., 2024). This constraint significantly reduces the informational complexity of the RBD. Thus, the two-letter approximation—which omits options 2 and 3 in the RBD—provides a reasonable simplification.

## 2, How to reconstruct the phylogenetic tree? [2][3]

In the four-letter representation of mutants, let $p_a$ denote the probability of a letter "*a*" (where $a \in \{0,1,2,3\}$) occurring in a sequence, and let $p_{ab}$ represent the joint probability of consecutive occurrence of letters "*a*" and "*b*". More generally, for any $n$-letter segment $\sigma = abc\ldots$, let $p_\sigma$ denote the joint probability of the bases in $\sigma$ occurring in the sequence. For joint probability calculations, all sequences are assumed to be circular. For any given $n$, the sum of joint probabilities over all segments $\sigma$ of length $n$ equals 1:

$$\sum_\sigma p_\sigma = 1,$$

where the summation extends over all $4^n$ possible segments of length $n$.

For two sequences $\Sigma$ and $\Sigma'$ with corresponding joint probability sets $\{p_\sigma\}$ and $\{p'_\sigma\}$, we define the $n$-mer distance between them based on joint probability differences as:

$$E_n(\Sigma, \Sigma') = \sum_\sigma |p_\sigma - p'_\sigma|. \qquad n=1,2,\ldots \tag{1}$$

The arguments of $E_n$ may be omitted when unambiguous. This $n$-mer distance is well-defined for unaligned sequences of unequal lengths. By iteratively applying the relation:

$$|p_\sigma - p'_\sigma| = |\sum_a (p_{\sigma a} - p'_{\sigma a})| \leq \sum_a |p_{\sigma a} - p'_{\sigma a}|, \tag{2}$$

where $\sigma$ is any $n$-letter segment and $\sigma a$ denotes an $(n+1)$-letter segment, it follows that:

$$E_{n+1} \geq E_n, \qquad n=1,2,\ldots \tag{3}$$

Three primary algorithms are commonly employed for phylogenetic tree inference: distance matrix methods, maximum parsimony analysis, and maximum likelihood estimation. Set $D_{ij}$ representing the n-mer distance between sequences $\Sigma_i$ and $\Sigma_j$. By applying the UPGMA algorithm to this matrix, we reconstructed an evolutionary tree for a specific $n$. We observed that the phylogenetic tree structure stabilizes after several iterations as $n$ increases, reaching a stable state when $n \geq 7$ (Luo and Lv, 2024).

## 3, How to generate new strains on the phylogenetic tree using a stochastic method? [2][3]

Once the phylogenetic tree is constructed from known viral mutant lineages, new strains can be generated and analyzed within this framework. We introduce a stochastic modeling approach for this process. Let set A comprise all mutated sites in updated virus strains, where a denotes the number of such sites. Each strain is represented as a two-letter sequence derived from A(a). To model stochastic mutations on the spike protein, let X denote a set of x sites subject to stochastic sampling. The union of sets A(a) and X(x) is defined as Z = A ∪ X, while their intersection, containing y sites, is denoted Y(y). Assuming that new strains arise from stochastic sampling within Z, each new strain is represented by a two-letter sequence from Z (len), where len = a + x − y. We reconstruct an evolutionary tree of mutants in set Z, which incorporates the predicted new strain. This methodology is termed the A–X model. Importantly, the model can consistently predict new strains, regardless of how sets A and X are distributed along the sequence or how set Y is incorporated. It ensures accurate predictions even with the continuous addition of new experimental data.

To demonstrate the feasibility of the A-X model in predicting new SARS-CoV-2 strains, we consider the example of 25 mutants listed in Supplementary Material 1, Table 1. Assume the evolutionary tree of these 25 mutants (set *A*) has two main branches, N and O. By performing stochastic sampling *S* times (e.g., $S = 10^5$), we generate $10^5$ predictions for new strains. The following observations are made:

1. The new strain may belong to either branch N or O, or lie outside both, suggesting the emergence of a new macro-lineage.

2. For a fixed *x*, different values of *y* can lead to distinct predictions regarding the position of the new strain on the tree. The position is determined by the pair (*x*, *y*). Thus, for a given (*x*, *y*), the predicted new strain is denoted as $New_{x.y}$.

3. When *x* is small, the predicted new strain typically falls within branch N or O. However, once *x* exceeds a certain threshold, an anomaly occurs, indicating the emergence of a new macro-lineage to which the new strain belongs.

## 4, How to predict new SARS-CoV-2 macro-lineages and what is the law for the generation of multiple macro-lineages? [3][5]

The emergence of new strains in phylogenetic trees is a stochastic process. As the scale of stochastic sampling expands, discernible statistical patterns begin to emerge. To broaden this sampling scope, we increased the size of *x* and randomized parameter *y* within the intersection set. This approach effectively enhances stochastic information, thereby facilitating new insights into the emergence of macro-lineages.

At a given time $t$, the emergence of new variants from set $A(a)$ can be predicted. In the A–X model, set $X$ is stochastically selected, resulting in the random appearance of new variants on the phylogenetic tree. Nevertheless, the generation of multiple macro-lineages and the consequent bifurcation of the tree topology into several major branches are inevitable.

For example, at $t = T_1$ (October 15, 2023; accession date of 25 mutants; see Table 1 in Appendix), we performed $10^5$ rounds of stochastic sampling for $x$ and 6 rounds of randomization for $y$. Throughout this process, the consistent emergence of macro-lineages N, O, and P was observed. The probability of a lineage (N, O, or P) for a given $x$ is calculated as the occurrence count of that lineage divided by the total number of variants ($10^5$). Specifically, the macro-lineage affiliation of a newly generated variant is determined by its position within the bifurcated phylogenetic tree. The results are shown in Fig. 1 (see Appendix), which plots the probability of the $j$-lineage ($j$ = N, O, P) against $x$.

A 99th percentile threshold was used in the calculation. We found that the macro-lineage assignment of a new variant depends statistically on x, with specific threshold values distinguishing different macro-lineages as follows:

$\text{New}_{x.y} \in$ N when $x \leq 18$; $\text{New}_{x.y} \in$ N or O when $18 < x \leq 29$; $\text{New}_{x.y} \in$ O when $29 < x \leq 36$; $\text{New}_{x.y} \in$ O or P when $36 < x \leq 69$; and $\text{New}_{x.y} \in$ P when $x > 69$.

At a later time point, $T_2$ (July 20, 2024), after incorporating 11 new mutants (listed in Table 2 , Appendix) into set $A$, the value of a increased from 104 to 128, covering 36 mutants in total. We repeated the stochastic sampling process with $10^5$ iterations for $x$ and 6 for $y$. The probabilities of each $j$-lineage ($j$ = N, O, P, Q) are presented in Fig. 2 (see Appendix).

Using the same method, we obtained the following threshold values of $x$ for the case of 36 mutants:

$\text{New}_{x.y} \in$ N when $x \leq 18$; $\text{New}_{x.y} \in$ N or O when $18 < x \leq 29$; $\text{New}_{x.y} \in$ O when $29 < x \leq 39$; $\text{New}_{x.y} \in$ O or P when $39 < x \leq 62$; $\text{New}_{x.y} \in$ P when $62 < x \leq 72$; $\text{New}_{x.y} \in$ P or Q when $72 < x \leq 99$ and $\text{New}_{x.y} \in$ Q when $x > 99$.

At the subsequent time point $T_3$ (January 4, 2025), after incorporating 70 variants from Table 3 into set $A$, the value of a increased to 153. The probabilities for each $j$-lineage ($j$ = N, O, P, Q) based on these 70 mutants are shown in Fig. 3 (see Appendix).. The corresponding threshold values of $x$ are defined as follows:

$\mathrm{New}_{x.y} \in$ N when $x \le 21$; $\mathrm{New}_{x.y} \in$ N or O when $21 < x \le 30$; $\mathrm{New}_{x.y} \in$ O when $30 < x \le 37$; $\mathrm{New}_{x.y} \in$ O or P when $37 < x \le 61$; $\mathrm{New}_{x.y} \in$ P when $61 < x \le 84$; $\mathrm{New}_{x.y} \in$ P or Q when $84 < x \le 112$ and $\mathrm{New}_{x.y} \in$ Q when $x > 112$.

***The law for generation of multiple macro-lineages***

Set $A(a)$ is an observable set dependent on time $t$. Set $X$ represents a set of assumed mutated sites, also time-dependent, which is used to predict new macro-lineages. We define $x_{\mathrm{dem1}}$ as the threshold value of $x$ at which a new macro-lineage emerges with a nonzero probability (greater than 1% in our calculations), and $x_{\mathrm{dem2}}$ as the threshold for its definitive occurrence. Both $x_{\mathrm{dem1}}$ and $x_{\mathrm{dem2}}$ are functions of time $t$.

At time $t_k$, let the value of $a$ be denoted as $a_k$, with examples $a_1 = 104$, $a_2 = 128$ and $a_3 = 153$. Denote $x_{\mathrm{dem1}}$ for macro-lineage $j$ (where $j$ = O, P, Q) at time $t_k$ as $x^{(j)}_{\mathrm{dem1}}(t_k)$, and similarly, denote $x_{\mathrm{dem2}}$ for macro-lineage $j$ at time $t_k$ as $x^{(j)}_{\mathrm{dem2}}(t_k)$. The values of $x^{(j)}_{\mathrm{dem1}}(t_k)$ and $x^{(j)}_{\mathrm{dem2}}(t_k)$ are shown in Figs. 1, 2 and 3, and summarized in Table 4 (see Appendix).

Table 4 outlines the occurrence pattern of novel macro-lineages: the O-lineage emerges at $x \approx 19$–22 and definitively occurs at $x \approx 30$–31; the P-lineage emerges at $x \approx 37$–40 and definitively occurs at $x \approx 62$–70; the Q-lineage emerges at $x \approx 73$–85 and definitively occurs at $x \approx 100$–113. The threshold values deduced across different time periods are highly consistent.

In conclusion, by conceptualizing stochastic mutation as a mechanism driving the emergence of new SARS-CoV-2 variants, we derived a principle governing macro-lineage transformation from N to P and then to Q. This approach offers a predictive framework for anticipating the future dominance of novel SARS-CoV-2 variants.

## 5, How to study the specific survival time for each macro-lineage of SARS-CoV-2 and predict the emergence time of new macro-lineage? [4][5]

According to the retrieval of SARS-CoV-2 variant data in Table S3, Table S5 presents the relationship between the number of mutated sites (NMS) in the spike protein and the initial global sample collection dates for each SARS-CoV-2 variant. We studied the relationship and discuss the possibly existing evolutionary clock defined by NMS in the spike protein. Although the mutation accumulation is affected by founder effects and immune escape fitness plateaus, we discover the number of mutated sites (NMS) in the spike protein for each variant can be arranged in chronological order of the first global sample collection dates. It means the approximate evolutionary clock does exist. Further, our analysis reveals that NMS($t$) represents a discontinuous function of time, corresponding to three distinct macro-lineages (Fig. 4A, Appendix). This figure is derived directly from the data in Table 5 (see Appendix).

***Linear regression of mutated sites vs sample collection date as an approximation***

We performed linear regression to analyze the relationship between the number of mutated sites and the initial sample collection date; the results are shown in Fig. 4B (see Appendix). The linear model effectively captures the increasing trend in mutated sites over viral evolution. The regression yielded an $R^2$ value of 0.92 and an RSE (Residual Standard Error) of 6.68. Furthermore, using linear regression theory, we estimated a slope of 1.163 per month (95% CI: ±0.076). Given the strong fit of this linear model, we will utilize the relationship between the number of mutated sites ($x$) and collection time ($t$) to predict the emergence of new viral lineages.

Considering the viral mutation accumulation rates are not uniform, if we replace the linear regression with a piecewise or segmented model, then Fig 4B will be replaced by Fig. 4A. the relationship between the number of mutated sites (NMS) and time ($t$) is discontinuous and exhibits stepwise changes at lineage transformation points. However, the slope of NMS increase differs between adjacent lineages. Typically, newer lineages show a lower rate of increase than older ones, offsetting the stepwise changes during lineage transformation. This compensatory mechanism accounts for the overall approximately linear relationship between NMS and emergence time. Our prediction for the new lineage relies on extrapolating this linear trend over an extended period.

***Prediction of the emergence time of the possible Q macro-lineage***

Based on the above analyses, we propose an algorithm for predicting the emergence timeline of macro-lineages. In the A–X model, the number of randomly mutated sites for a variant is denoted as $x$. Notably, NMS($t$), which quantifies the accumulation of mutated sites over time, is intrinsically related to $x$. Using the data from Fig. 4—which shows the relationship between the probability of macro-lineage emergence (PML) and $x$, as well as the time dependence of NMS($t$)—we predict the emergence time of the new macro-lineage Q. By extrapolating the linear trend over a longer period and applying linear regression, we estimate that the number of mutated sites will reach NMS =73.4 ±2.6 to 85 ± 3 between the 62nd and 72nd months, and 100 ± 4 to 113 ± 5 between the 85th and 96th months, starting from December 2019. Combining these values with the thresholds $x_{\text{dem1}}$ = 73-85 (indicating the initial emergence of Q) and $x_{\text{dem2}}$ = 100-113 (indicating a major outbreak of Q), we preliminarily predict that macro-lineage Q will emerge around February to December 2025, and may experience a strong outbreak approximately 22 to 24 months later.

## 6, Is there any phenomenological model to analyze the spreading dynamics of COVID-19? [1]

A mathematical model is proposed to analyze the spreading dynamics of COVID-19. By using the parameters of the model, namely the basic reproduction number ($R_0$) and the attenuation constant ($k$), the daily number of infections (DNI) and the cumulative number of infections (CNI) over time ($m$) are deduced and shown to be in good agreement with experimental data. This model effectively addresses three key issues: (1) inferring the conditions under which virus infections die out for a specific strain given $R_0$; (2) explaining the occurrence of second waves of infection and developing preventive measures; and (3) understanding the competitive spread of two viruses within a region and devising control strategies. The findings highlight the potential of this simple mathematical framework in comprehensively addressing these challenges. The theoretical insights derived from this model can guide the evaluation of infection wave severity and the formulation of effective strategies for controlling and mitigating epidemic outbreaks.

# Appendix

## Part 1 Mutation data of SARS-CoV-2 Variants

**Table 1 Partial table of 25 variants and their mutation sites.** A total of 25 variants from the mutation reports provided by outbreak.info (https://outbreak.info/, accessed on 15 October 2023)

| macro-lineage | No. | variant | NMS | mutated sites |
|---|---|---|---|---|
| N-lineage | 1 | P.2(Zeta) | 3 | **484**,614,1176 |
| | 2 | B.1.1.7(Alpha) | 10 | 69,70,144,**501**,570,614,681,716,982,1118 |
| | 3 | B.1.429(Epsilon) | 4 | 13,152,**452**,614 |
| | 4 | B.1.351(Beta) | 10 | 80,215,241,242,243,**417**,**484**,**501**,614,701 |
| | 5 | B.1.617.2(Delta) | 9 | 19,156,157,158,**452**,**478**,614,681,950 |
| | 6 | P.1(Gamma) | 12 | 18,20,26,138,190,**417**,**484**,**501**,614,655,1027,1176 |
| | 7 | B.1.617.1(Kappa) | 5 | **452**,**484**,614,681,1071 |
| | 8 | B.1.621(Mu) | 9 | 95,144,145,**346**,**484**,**501**,614,681,950 |
| | 9 | C.37(Lambda) | 14 | 75,76,246,247,248,249,250,251,252,253,**452**,**490**,614,859 |
| | 10 | B.1.526(Iota) | 4 | 5,95,253,614 |
| | 11 | B.1.525(Eta) | 9 | 52,67,69,70,144,**484**,614,677,888 |
| | 12 | P.3(Theta) | 7 | **484**,**501**,614,681,1092,1101,1176 |
| O-lineage | 13 | BA.1 | 33 | 67,69,70,95,142,143,144,145,211,212,**339,371,373,375,477,478,484,493,496,498,501,505**,547,614,655,679,681,764,796,856,954,969,981 |
| | 14 | BA.2 | 31 | 19,24,25,26,27,142,213,**339,371,373,375,376,405,408,417,440,477,478,484,493,498,501,505**,614,655,679,681,764,796,954,969 |
| | 15 | BA.2.12.1 | 33 | 19,24,25,26,27,142,213,**339,371,373,375,376,405,408,417,440,452,477,478,484,493,498,501,505**,614,655,679,681,704,764,796,954,969 |
| | 16 | BA.5 | 34 | 19,24,25,26,27,69,70,142,213,**339,371,373,375,376,405,408,417,440,452,477,478,484,486,498,501,505**,614,655,679,681,764,796,954,969 |
| | 17 | BA.4.1 | 35 | 3,19,24,25,26,27,69,70,142,213,**339,371,373,375,376,405,408,417,440,452,477,478,484,486,498,501,505**,614,655,679,681,764,796,954,969 |
| | 18 | BQ.1.1 | 37 | 19,24,25,26,27,69,70,142,213,**339,346,371,373,375,376,405,408,417,440,444,452,460,477,478,484,486,498,501,505**,614,655,679,681,764,796,954,969 |
| | 19 | BA.2.75 | 30 | 19,24,210,213,257,**339**,**371**,**373**,**375**,**376**,**405**,**408**,**417**,**440**,**446**,**460**,**477**,**478**,**484**,**498**,**501**,**505**,614,655,679,681,764,796,954,969 |
| | 20 | BF.7 | 35 | 19,24,25,26,27,69,70,142,213,**339,346,371,373,375,376,405,408,417,440,452,477,478,484,486,498,501,505**,614,655,679,681,764,796,954,969 |
| | 21 | CH.1.1 | 41 | 19,24,25,26,27,142,147,152,157,210,213,257,**339,346,371,373,375,376,405,408,417,440,444,446,452,460,477,478,484,486,498,501,505**,614,655,679,681,764,796,954,969 |
| | 22 | XBB.1.5 | 42 | 19,24,25,26,27,83,142,144,146,183,213,252,**339,346,368,371,373,375,376,405,408,417,440,445,446,460,477,478,484,486,490,498,501,505**,614,655,679,681,764,796,954,969 |
| | 23 | XBB.1.16 | 43 | 19,24,25,26,27,83,142,144,146,180,183,213,252,**339,346,368** |
| | | | | **,371,373,375,376,405,408,417,440,445,446,460,477,478,484,486,490,498,501,505**,614,655,679,681,764,796,954,969 |
| | 24 | EG.1 | 43 | 19,24,25,26,27,83,142,144,146,183,213,252,**339,346,368,371,373,375,376,405,408,417,440,445,446,460,477,478,484,486,490,498,501,505**,613,614,655,679,681,764,796,954,969 |
| | 25 | EG.5.1 | 44 | 19,24,25,26,27,52,83,142,144,146,183,213,252,**339,346,368,371,373,375,376,405,408,417,440,445,446,456,460,477,478,484,486,490,498,501,505**,614,655,679,681,764,796,954,969 |
| total | | | 104 | 3, 5, 13, 18, 19, 20, 24, 25, 26, 27, 52, 67, 69, 70, 75, 76, 80, 83, 95, 138, 142, 143, 144, 145, 146, 147, 152, 156, 157, 158, 180, 183, 190, 210, 211, 212, 213, 215, 241, 242, 243, 246, 247, 248, 249, 250, 251, 252, 253, 257, **339, 346, 368, 371, 373, 375, 376, 405, 408, 417, 440, 444, 445, 446, 452, 456, 460, 477, 478, 484, 486, 490, 493, 496, 498, 501, 505**, 547, 570, 613, 614, 655, 677, 679, 681, 701, 704, 716, 764, 796, 856, 859, 888, 950, 954, 969, 981, 982, 1027, 1071, 1092, 1101, 1118, 1176 |

NMS: number of mutated sites; Mutated sites are considered when they occur in at least 75% of the SARS-CoV-2 lineage sequences. Mutations in the spike protein's receptor-binding domain (RBD) are indicated in bold.

**Table 2 Partial table of 36 variants and their mutation sites.** A total of 36 variants from the mutation reports provided by outbreak.info (https://outbreak.info/, accessed on 20 July 2024)

| macro-lineage | No. | variant | NMS | mutated sites |
|---|---|---|---|---|
| N-lineage | 1 | P.2(Zeta) | 3 | **484**,614,1176 |
| | 2 | B.1.1.7(Alpha) | 10 | 69,70,144,**501**,570,614,681,716,982,1118 |
| | 3 | B.1.429(Epsilon) | 4 | 13,152,**452**,614 |
| | 4 | B.1.351(Beta) | 10 | 80,215,241,242,243,**417**,**484**,**501**,614,701 |
| | 5 | B.1.617.2(Delta) | 9 | 19,156,157,158,**452**,**478**,614,681,950 |
| | 6 | P.1(Gamma) | 12 | 18,20,26,138,190,**417**,**484**,**501**,614,655,1027,1176 |
| | 7 | B.1.617.1(Kappa) | 5 | **452**,**484**,614,681,1071 |
| | 8 | B.1.621(Mu) | 9 | 95,144,145,**346**,**484**,**501**,614,681,950 |
| | 9 | C.37(Lambda) | 14 | 75,76,246,247,248,249,250,251,252,253,**452**,**490**,614,859 |
| | 10 | B.1.526(Iota) | 4 | 5,95,253,614 |
| | 11 | B.1.525(Eta) | 9 | 52,67,69,70,144,**484**,614,677,888 |
| | 12 | P.3(Theta) | 7 | **484**,**501**,614,681,1092,1101,1176 |
| O-lineage | 13 | BA.1 | 33 | 67,69,70,95,142,143,144,145,211,212,**339,371,373,375,477,478,484,493,496,498,501,505**,547,614,655,679,681,764,796,856,954,969,981 |
| | 14 | BA.2 | 31 | 19,24,25,26,27,142,213,**339,371,373,375,376,405,408,417,440,477,478,484,493,498,501,505**,614,655,679,681,764,796,954,969 |
| | 15 | BA.2.12.1 | 33 | 19,24,25,26,27,142,213,**339,371,373,375,376,405,408,417,440,452,477,478,484,493,498,501,505**,614,655,679,681,704,764,796,954,969 |
| | 16 | BA.5 | 34 | 19,24,25,26,27,69,70,142,213,**339,371,373,375,376,405,408,417,440,452,477,478,484,486,498,501,505**,614,655,679,681,764,796,954,969 |
| | 17 | BA.4.1 | 35 | 3,19,24,25,26,27,69,70,142,213,**339,371,373,375,376,405,408,417,440,452,477,478,484,486,498,501,505**,614,655,679,681,764,796,954,969 |
| | 18 | BQ.1.1 | 37 | 19,24,25,26,27,69,70,142,213,**339,346,371,373,375,376,405,408,417,440,444,452,460,477,478,484,486,498,501,505**,614,655,679,681,764,796,954,969 |
| | 19 | BA.2.75 | 30 | 19,24,210,213,257,**339,371,373,375,376,405,408,417,440,446,460,477,478,484,498,501,505**,614,655,679,681,764,796,954,969 |
| | 20 | BF.7 | 35 | 19,24,25,26,27,69,70,142,213,**339,346,371,373,375,376,405,408,417,440,452,477,478,484,486,498,501,505**,614,655,679,681,764,796,954,969 |
| | 21 | CH.1.1 | 41 | 19,24,25,26,27,142,147,152,157,210,213,257,**339,346,371,373,375,376,405,408,417,440,444,446,452,460,477,478,484,486,498,501,505**,614,655,679,681,764,796,954,969 |
| | 22 | XBB.1.5 | 42 | 19,24,25,26,27,83,142,144,146,183,213,252,**339,346,368,371,373,375,376,405,408,417,440,445,446,460,477,478,484,486,490,498,501,505**,614,655,679,681,764,796,954,969 |
| | 23 | XBB.1.16 | 43 | 19,24,25,26,27,83,142,144,146,180,183,213,252,**339,346,368,371,373,375,376,405,408,417,440,445,446,460,477,478,484,486,490,498,501,505**,614,655,679,681,764,796,954,969 |
| | 24 | EG.1 | 43 | 19,24,25,26,27,83,142,144,146,183,213,252,**339,346,368,371,373,375,376,405,408,417,440,445,446,460,477,478,484,486,490,498,501,505**,613,614,655,679,681,764,796,954,969 |
| | 25 | EG.5.1 | 44 | 19,24,25,26,27,52,83,142,144,146,183,213,252,**339,346,368,371,373,375,376,405,408,417,440,445,446,456,460,477,478,484,486,490,498,501,505**,614,655,679,681,764,796,954,969 |
| | 26 | HV.1 | 46 | 19,24,25,26,27,52,83,142,144,146,157,183,213,252,**339,346,368,371,373,375,376,405,408,417,440,445,446,452,456,460,477,478,484,486,490,498,501,505**,614,655,679,681,764,796, |
| | | | | 954,969 |
| P-lineage | 27 | JN.1 | 60 | 19,21,24,25,26,27,50,69,70,127,142,144,157,158,211,212,213,216,245,264,**332,339,356,371,373,375,376,403,405,408,417,440,445,446,450,452,455,460,477,478,481,483,484,486,498,501,505**,554,570,614,621,655,679,681,764,796,939,954,969,1143 |
| | 28 | BA.2.86.1 | 59 | 19,21,24,25,26,27,50,69,70,127,142,144,157,158,211,212,213,216,245,264,**332,339,356,371,373,375,376,403,405,408,417,440,445,446,450,452,460,477,478,481,483,484,486,498,501,505**,554,570,614,621,655,679,681,764,796,939,954,969,1143 |
| | 29 | BA.2.86 | 58 | 19,21,24,25,26,27,50,69,70,127,142,144,157,158,211,212,213,216,245,264,**332,339,356,371,373,375,376,403,405,408,417,440,445,446,450,452,460,477,478,481,484,486,498,501,505**,554,570,614,621,655,679,681,764,796,939,954,969,1143 |
| | 30 | JN.1.7 | 62 | 19,21,24,25,26,27,50,69,70,127,142,144,157,158,211,212,213,216,245,264,**332,339,356,371,373,375,376,403,405,408,417,440,445,446,450,452,455,460,477,478,481,483,484,486,498,501,505**,554,570,572,614,621,655,679,681,764,796,939,954,969,1143,1150 |
| | 31 | KP.3.1.1 | 64 | 19,21,24,25,26,27,31,50,69,70,127,142,144,157,158,211,212,213,216,245,264,**332,339,356,371,373,375,376,403,405,408,417,440,445,446,450,452,455,456,460,477,478,481,483,484,486,493,498,501,505**,554,570,614,621,655,679,681,764,796,939,954,969,1104,1143 |
| | 32 | KP.2 | 59 | 19,21,50,69,70,127,142,144,157,158,211,212,213,216,245,264,**332,339,346,356,371,373,375,376,403,405,408,417,440,445,446,450,452,455,456,460,477,478,481,483,484,486,498,501,505**,554,570,614,621,655,679,681,764,796,939,954,969,1104,1143 |
| | 33 | JN.1.37 | 61 | 19,21,24,25,26,27,50,69,70,127,142,144,157,158,211,212,213,216,245,264,**332,339,356,371,373,375,376,403,405,408,417,440,445,446,450,452,455,460,477,478,481,483,484,486,498,501,505**,554,570,614,621,655,679,680,681,764,796,939,954,969,1143 |
| | 34 | XDQ.1 | 55 | 19,21,24,25,26,27,50,69,70,127,142,144,157,158,211,212,213,216,245,264,**332,339,356,371,373,375,376,403,405,408,417,440,445,446,450,452,460,475,483,484,486,498,501,505**,554,570,614,621,655,679,681,764,796,954,969 |
| | 35 | LB.1 | 64 | 19,21,24,25,26,27,31,50,69,70,127,142,144,157,158,183,211,212,213,216,245,264,**332,339,346,356,371,373,375,376,403,405,408,417,440,445,446,450,452,455,456,460,477,478,481,483,484,486,498,501,505**,554,570,614,621,655,679,681,764,796,939,954,969,1143 |
| | 36 | KP.1 | 63 | 19,21,24,25,26,27,50,69,70,127,142,144,157,158,211,212,213,216,245,264,**332,339,356,371,373,375,376,403,405,408,417,440,445,446,450,452,455,456,460,477,478,481,483,484,486,498,501,505**,554,570,614,621,655,679,681,764,796,939,954,969,1086,1104,1143 |
| total | | | 128 | 3,5,13,18,19,20,21,24,25,26,27,31,50,52,67,69,70,75,76,80,83,95,127,138,142,143,144,145,146,147,152,156,157,158,180,183,190,210,211,212,213,215,216,241,242,243,245,246,247,248,249,250,251,252,253,257,264,**332,339,346,356,368,371,373,375,376,403,405,408,417,440,444,445,446,450,452,455,456,460,475,477,478,481,483,484,486,490,493,496,498,501,505**,547,554,570,572,613,614,621,655,677,679,680,681,701,704,716,764,796,856,859,888,939,950,954,969,981,982,1027,1071,1086,1092,1101,1104,1118,1143,1150,1176 |

NMS: number of mutated sites; Mutated sites are considered when they occur in at least 75% of the SARS-CoV-2 lineage sequences. Mutations in the spike protein's receptor-binding domain (RBD) are indicated in bold.

**Table 3 General table of SARS-CoV-2 70 variants and their mutation sites**. A total of 70 variants from the mutation reports provided by outbreak.info (https://outbreak.info/, accessed on 4 January 2025)

| **macro-lineage** | **No.** | **variant** | **NMS** | **mutated sites** |
|---|---|---|---|---|
| N-lineage | 1 | B.1 | 1 | 614 |
| | 2 | B.1.177 | 2 | 222,614 |
| | 3 | B.1.1.176 | 3 | 20,**357**,614 |
| | 4 | D.2 | 2 | **477**,614 |
| | 5 | P.2(Zeta) | 3 | **484**,614,1176 |
| | 6 | B.1.1.7(Alpha) | 10 | 69,70,144,**501**,570,614,681,716,982,1118 |
| | 7 | B.1.429(Epsilon) | 4 | 13,152,**452**,614 |
| | 8 | B.1.351(Beta) | 10 | 80,215,241,242,243,**417**,**484**,**501**,614,701 |
| | 9 | B.1.617.2(Delta) | 9 | 19,156,157,158,**452**,**478**,614,681,950 |
| | 10 | P.1(Gamma) | 12 | 18,20,26,138,190,**417**,**484**,**501**,614,655,1027,1176 |
| | 11 | B.1.617.1(Kappa) | 5 | **452**,**484**,614,681,1071 |
| | 12 | B.1.621(Mu) | 9 | 95,144,145,**346**,**484**,**501**,614,681,950 |
| | 13 | R.1 | 4 | 152,**484**,614,769 |
| | 14 | C.37(Lambda) | 14 | 75,76,246,247,248,249,250,251,252,253,**452**,**490**,614,859 |
| | 15 | B.1.526(Iota) | 4 | 5,95,253,614 |
| | 16 | B.1.525(Eta) | 9 | 52,67,69,70,144,**484**,614,677,888 |
| | 17 | C.36.3 | 9 | 12,69,70,152,**346**,**452**,614,677,899 |
| | 18 | P.3(Theta) | 7 | **484**,**501**,614,681,1092,1101,1176 |
| | 19 | AZ.2 | 6 | 95,144,**484**,614,681,796 |
| | 20 | AV.1 | 10 | 80,95,142,144,**439**,**484**,614,681,1130,1139 |
| | 21 | B.1.1.529 | 7 | **373**,**478**,614,655,679,681,954 |
| | 22 | C.1.2 | 15 | 9,136,144,190,215,243,244,**449**,**484**,**501**,614,655,679,716,859 |
| O-lineage | 23 | BA.1 | 33 | 67,69,70,95,142,143,144,145,211,212,**339,371,373,375,477,478,484,493,496,498,501,505**,547,614,655,679,681,764,796,856,954,969,981 |
| | 24 | BA.1.1 | 35 | 67,69,70,95,142,143,144,145,211,212,**339,346,371,373,375,446,477,478,484,493,496,498,501,505**,547,614,655,679,681,764,796,856,954,969,981 |
| | 25 | BA.2 | 31 | 19,24,25,26,27,142,213,**339,371,373,375,376,405,408,417,440,477,478,484,493,498,501,505**,614,655,679,681,764,796,954,969 |
| | 26 | BA.2.12.1 | 33 | 19,24,25,26,27,142,213,**339,371,373,375,376,405,408,417,440,452,477,478,484,493,498,501,505**,614,655,679,681,704,764,796,954,969 |
| | 27 | BA.2.65 | 31 | 19,24,25,26,27,142,213,**339,371,373,375,376,405,408,417,440,477,478,484,493,498,501,505**,614,655,679,681,764,796,954,969 |
| | 28 | BA.1.1.15 | 37 | 67,69,70,95,142,143,144,145,211,212,**339,346,371,373,375,417,440,446,477,478,484,493,496,498,501,505**,547,614,655,679,681,764,796,856,954,969,981 |
| | 29 | BA.5 | 34 | 19,24,25,26,27,69,70,142,213,**339,371,373,375,376,405,408,417,440,452,477,478,484,486,498,501,505**,614,655,679,681,764,796,954,969 |
| | 30 | BA.4.1 | 35 | 3,19,24,25,26,27,69,70,142,213,**339,371,373,375,376,405,408,417,440,452,477,478,484,486,498,501,505**,614,655,679,681,764,796,954,969 |
| | 31 | BQ.1.1 | 37 | 19,24,25,26,27,69,70,142,213,**339,346,371,373,375,376,405,408,417,440,444,452,460,477,478,484,486,498,501,505**,614,655,679,681,764,796,954,969 |
| | 32 | BA.2.75 | 30 | 19,24,210,213,257,**339,371,373,375,376,405,408,417,440,446,460,477,478,484,498,501,505**,614,655,679,681,764,796,954,969 |
| | 33 | BF.5 | 35 | 19,24,25,26,27,69,70,142,213,**339,371,373,375,376,405,408,417,440,452,477,478,484,486,498,501,505**,614,655,679,681,764,796,954,969,1020 |
| | 34 | BF.7 | 35 | 19,24,25,26,27,69,70,142,213,**339,346,371,373,375,376,405,408,417,440,452,477,478,484,486,498,501,505**,614,655,679,681,764,796,954,969 |
| | 35 | BN.1.2 | 40 | 19,24,25,26,27,142,147,152,157,210,213,257,**339,346,356,371,373,375,376,405,408,417,440,446,460,477,478,484,490,498,501,505**,614,655,679,681,764,796,954,969 |
| | 36 | CH.1.1 | 41 | 19,24,25,26,27,142,147,152,157,210,213,257,**339,346,371,373,375,376,405,408,417,440,444,446,452,460,477,478,484,486,498,501,505**,614,655,679,681,764,796,954,969 |
| | 37 | XBB.1.5 | 42 | 19,24,25,26,27,83,142,144,146,183,213,252,**339,346,368,371,373,375,376,405,408,417,440,445,446,460,477,478,484,486,490,498,501,505**,614,655,679,681,764,796,954,969 |
| | 38 | BM.4.1.1 | 39 | 19,24,25,26,27,142,147,152,157,210,213,257,**339,346,371,373,375,376,405,408,417,440,446,460,477,478,484,486,498,501,505**,614,655,679,681,764,796,954,969 |
| | 39 | XBC.1 | 39 | 19,25,142,144,156,157,158,209,212,215,222,243,244,**371,373,375,376,405,408,417,440,446,452,477,478,484,486,498,501,505**,614,655,679,681,703,764,796,954,969 |
| | 40 | CH.1.1.1 | 42 | 19,24,25,26,27,142,147,152,157,185,210,213,257,**339,346,371,373,375,376,405,408,417,440,444,446,452,460,477,478,484,486,498,501,505**,614,655,679,681,764,796,954,969 |
| | 41 | XBB.1.16 | 43 | 19,24,25,26,27,83,142,144,146,180,183,213,252,**339,346,368,371,373,375,376,405,408,417,440,445,446,460,477,478,484,486,490,498,501,505**,614,655,679,681,764,796,954,969 |
| | 42 | EG.1 | 43 | 19,24,25,26,27,83,142,144,146,183,213,252,**339,346,368,371,373,375,376,405,408,417,440,445,446,460,477,478,484,486,490,498,501,505**,613,614,655,679,681,764,796,954,969 |
| | 43 | HV.1 | 46 | 19,24,25,26,27,52,83,142,144,146,157,183,213,252,**339,346,368,371,373,375,376,405,408,417,440,445,446,452,456,460,477,478,484,486,490,498,501,505**,614,655,679,681,764,796,954,969 |
| | 44 | HK.3 | 45 | 19,24,25,26,27,52,83,142,144,146,183,213,252,**339,346,368,371,373,375,376,405,408,417,440,445,446,455,456,460,477,478,484,486,490,498,501,505**,614,655,679,681,764,796,954,969 |
| | 45 | EG.5.1 | 44 | 19,24,25,26,27,52,83,142,144,146,183,213,252,**339,346,368,371,373,375,376,405,408,417,440,445,446,456,460,477,478,484,486,490,498,501,505**,614,655,679,681,764,796,954,969 |
| | 46 | DV.7.1 | 45 | 19,24,25,26,27,142,147,152,157,185,210,213,257,**339,346,371,373,375,376,405,408,417,440,444,446,452,455,456,460,477,478,484,486,498,501,505**,614,655,679,681,764,796,858,954,969 |
| P-lineage | 47 | JN.1 | 60 | 19,21,24,25,26,27,50,69,70,127,142,144,157,158,211,212,213,216,245,264,**332,339,356,371,373,375,376,403,405,408,417,440,445,446,450,452,455,460,477,478,481,483,484,486,498,501,505**,554,570,614,621,655,679,681,764,796,939,954,969,1143 |
| | 48 | BA.2.86.1 | 59 | 19,21,24,25,26,27,50,69,70,127,142,144,157,158,211,212,213,216,245,264,**332,339,356,371,373,375,376,403,405,408,417,440,445,446,450,452,460,477,478,481,483,484,486,498,501,505**,554,570,614,621,655,679,681,764,796,939,954,969,1143 |
| | 49 | BA.2.86 | 58 | 19,21,24,25,26,27,50,69,70,127,142,144,157,158,211,212,213,216,245,264,**332,339,356,371,373,375,376,403,405,408,417,440,445,446,450,452,460,477,478,481,484,486,498,501,505**,554,570,614,621,655,679,681,764,796,939,954,969,1143 |
| | 50 | JN.2 | 59 | 19,21,24,25,26,27,50,69,70,127,142,144,157,158,211,212,21 |

| | | | |
|---|---|---|---|
| | | | 3,216,245,264,**332,339,356,371,373,375,376,403,405,408,417,440,445,446,450,452,460,477,478,481,483,484,486,498,501,505**,554,570,614,621,655,679,681,764,796,939,954,969,1143 |
| 51 | JN.1.7 | 62 | 19,21,24,25,26,27,50,69,70,127,142,144,157,158,211,212,213,216,245,264,**332,339,356,371,373,375,376,403,405,408,417,440,445,446,450,452,455,460,477,478,481,483,484,486,498,501,505**,554,570,572,614,621,655,679,681,764,796,939,954,969,1143,1150 |
| 52 | JN.1.18 | 57 | 19,21,50,69,70,127,142,144,157,158,211,212,213,216,245,264,**332,339,346,356,371,373,375,376,403,405,408,417,440,445,446,450,452,455,460,477,478,481,483,484,486,498,501,505**,554,570,614,621,655,679,681,764,796,939,954,969,1143 |
| 53 | JN.1.11.1 | 62 | 19,21,24,25,26,27,50,69,70,127,142,144,157,158,211,212,213,216,245,264,**332,339,356,371,373,375,376,403,405,408,417,440,445,446,450,452,455,456,460,477,478,481,483,484,486,498,501,505**,554,570,614,621,655,679,681,764,796,939,954,969,1104,1143 |
| 54 | KP.3.1.1 | 64 | 19,21,24,25,26,27,31,50,69,70,127,142,144,157,158,211,212,213,216,245,264,**332,339,356,371,373,375,376,403,405,408,417,440,445,446,450,452,455,456,460,477,478,481,483,484,486,493,498,501,505**,554,570,614,621,655,679,681,764,796,939,954,969,1104,1143 |
| 55 | KP.2 | 59 | 19,21,50,69,70,127,142,144,157,158,211,212,213,216,245,264,**332,339,346,356,371,373,375,376,403,405,408,417,440,445,446,450,452,455,456,460,477,478,481,483,484,486,498,501,505**,554,570,614,621,655,679,681,764,796,939,954,969,1104,1143 |
| 56 | JN.1.37 | 61 | 19,21,24,25,26,27,50,69,70,127,142,144,157,158,211,212,213,216,245,264,**332,339,356,371,373,375,376,403,405,408,417,440,445,446,450,452,455,460,477,478,481,483,484,486,498,501,505**,554,570,614,621,655,679,680,681,764,796,939,954,969,1143 |
| 57 | XEB | 61 | 19,21,24,25,26,27,50,69,70,127,142,144,157,158,211,212,213,216,245,264,**332,339,356,371,373,375,376,403,405,408,417,440,445,446,450,452,455,460,477,478,481,483,484,486,498,501,505**,554,570,614,621,655,679,681,764,796,939,954,969,1143,1174 |
| 58 | XDQ.1 | 55 | 19,21,24,25,26,27,50,69,70,127,142,144,157,158,211,212,213,216,245,264,**332,339,356,371,373,375,376,403,405,408,417,440,445,446,450,452,460,475,483,484,486,498,501,505**,554,570,614,621,655,679,681,764,796,954,969 |
| 59 | KP.3 | 63 | 19,21,24,25,26,27,50,69,70,127,142,144,157,158,211,212,213,216,245,264,**332,339,356,371,373,375,376,403,405,408,417,440,445,446,450,452,455,456,460,477,478,481,483,484,486,493,498,501,505**,554,570,614,621,655,679,681,764,796,939,954,969,1104,1143 |
| 60 | LB.1 | 64 | 19,21,24,25,26,27,31,50,69,70,127,142,144,157,158,183,211,212,213,216,245,264,**332,339,346,356,371,373,375,376,403,405,408,417,440,445,446,450,452,455,456,460,477,478,481,483,484,486,498,501,505**,554,570,614,621,655,679,681,764,796,939,954,969,1143 |
| 61 | KP.1 | 63 | 19,21,24,25,26,27,50,69,70,127,142,144,157,158,211,212,213,216,245,264,**332,339,356,371,373,375,376,403,405,408,417,440,445,446,450,452,455,456,460,477,478,481,483,484,486,498,501,505**,554,570,614,621,655,679,681,764,796,939,954,969,1086,1104,1143 |
| 62 | KS.1 | 58 | 50,59,69,70,127,142,144,157,158,211,212,213,216,245,264,**332,339,346,356,371,373,375,376,403,405,408,417,440,445,446,450,452,455,456,460,477,478,481,483,484,486,498,501,505**,554,570,614,621,655,679,681,764,796,939,954,969,1087,1143 |
| 63 | KP.1.1.3 | 65 | 19,21,24,25,26,27,31,50,69,70,127,142,144,157,158,211,212,213,216,245,264,**332,339,346,356,371,373,375,376,403,405,408,417,440,445,446,450,452,455,456,460,477,478,481,483,484,486,498,501,505**,554,570,614,621,655,679,681,764,796, |

| | | | | |
|---|---|---|---|---|
| | | | | 939,954,969,1086,1104,1143 |
| | 64 | XDV.1 | 56 | 19,21,50,69,70,127,142,144,157,158,211,212,213,216,245,264,**332,339,356,371,373,375,376,403,405,408,417,440,445,446,450,452,455,456,460,477,478,481,484,486,498,501,505**,554,570,614,621,655,679,681,764,796,939,954,969,1143 |
| | 65 | LP.1 | 66 | 19,21,24,25,26,27,31,50,69,70,127,142,144,157,158,211,212,213,216,245,264,**332,339,346,356,371,373,375,376,403,405,408,417,440,445,446,450,452,455,456,460,477,478,481,483,484,486,498,501,505**,554,570,614,621,655,679,681,764,796,939,954,969,1086,1104,1143,1229 |
| | 66 | XED | 64 | 19,21,24,25,26,27,31,50,69,70,127,142,144,157,158,211,212,213,216,245,264,**332,339,346,356,371,373,375,376,403,405,408,417,440,445,446,450,452,455,456,460,477,478,481,483,484,486,498,501,505**,554,570,614,621,655,679,681,764,796,939,954,969,1143,1263 |
| | 67 | XEC | 65 | 19,21,22,24,25,26,27,50,59,69,70,127,142,144,157,158,211,212,213,216,245,264,**332,339,356,371,373,375,376,403,405,408,417,440,445,446,450,452,455,456,460,477,478,481,483,484,486,493,498,501,505**,554,570,614,621,655,679,681,764,796,939,954,969,1104,1143 |
| | 68 | LF.7 | 67 | 19,21,22,24,25,26,27,31,50,69,70,127,142,144,157,158,182,190,211,212,213,216,245,264,**332,339,346,356,371,373,375,376,403,405,408,417,440,444,445,446,450,452,455,456,460,477,478,481,483,484,486,498,501,505**,554,570,614,621,655,679,681,764,796,939,954,969,1143 |
| | 69 | LF.7.1.2 | 68 | 19,21,22,24,25,26,27,31,50,69,70,127,142,144,157,158,182,190,211,212,213,216,245,264,**332,339,346,356,371,373,375,376,403,405,408,417,440,444,445,446,450,452,455,456,460,475,477,478,481,483,484,486,498,501,505**,554,570,614,621,655,679,681,764,796,939,954,969,1143 |
| | 70 | LP.8.1 | 68 | 19,21,24,25,26,27,31,50,69,70,127,142,144,157,158,186,190,211,212,213,216,245,264,**332,339,346,356,371,373,375,376,403,405,408,417,440,445,446,450,452,455,456,460,477,478,481,483,484,486,493,498,501,505**,554,570,614,621,655,679,681,764,796,939,954,969,1086,1104,1143 |
| total | | | 153 | 3,5,9,12,13,18,19,20,21,22,24,25,26,27,31,50,52,59,67,69,70,75,76,80,83,95,127,136,138,142,143,144,145,146,147,152,156,157,158,180,182,183,185,186,190,209,210,211,212,213,215,216,222,241,242,243,244,245,246,247,248,249,250,251,252,253,257,264,**332,339,346,356,357,368,371,373,375,376,403,405,408,417,439,440,444,445,446,449,450,452,455,456,460,475,477,478,481,483,484,486,490,493,496,498,501,505**,547,554,570,572,613,614,621,655,677,679,680,681,701,703,704,716,764,769,796,856,858,859,888,899,939,950,954,969,981,982,1020,1027,1071,1086,1087,1092,1101,1104,1118,1130,1139,1143,1150,1174,1176,1229,1263 |

NMS: number of mutated sites; Mutated sites are considered when they occur in at least 75% of the SARS-CoV-2 lineage sequences. Mutations in the spike protein's receptor-binding domain (RBD) are indicated in bold.

**Note:** 70 variants written in red are taken from the mutation reports provided by non- VOC samples in outbreak.info. It leads to the total number of mutated sites from 147 (in original case of 63 variants) increased to 153.

# Part 2 Generation of multiple macro-lineages

A-X model : Let $A$ represent the set of all mutated sites in the updated virus strains, with $a$ denoting the number of these sites. Considering stochastic mutations on the spike protein, let $X$ be the set of $x$ sites involved in the stochastic sampling. The union of sets $A(a)$ and $X(x)$ is denoted as $Z$, where $Z=A\cup X$. The intersection of sets $A$ and $X$, containing $y$ sites, is represented as $Y(y)$. Assuming the new strain arises from stochastic sampling within the union set $Z$(len), where len=$a$+$x$-$y$.

At a given time $t$, we can predict the emergence of new variants from set $A(a)$, the selection of set $X$ is stochastic, leading to the random appearance of new variants on the phylogenetic tree. However, we found the generation of multiple macro-lineages and the subsequent bifurcation of the tree topology into several major branches are inevitable.

For instance, at $t = T_1$ (October 15, 2023, accession date of 25 mutants, see Table 1), we performed $10^5$ rounds of stochastic sampling for $x$ and 6 rounds of randomization for $y$. During this process, we consistently observed the inevitable emergence of macro-lineages N, O, and P. The probability of a lineage (N, O, or P) for a given $x$ is calculated by dividing the occurrence count of each lineage by the total number of variants ($10^5$). Specifically, the macro-lineage to which a newly generated variant belongs is determined by its occurrence in the bifurcation of the phylogenetic tree. The results are illustrated in Fig. 1, where the probability of the $j$-lineage ($j$=N, O, or P) is plotted against $x$.

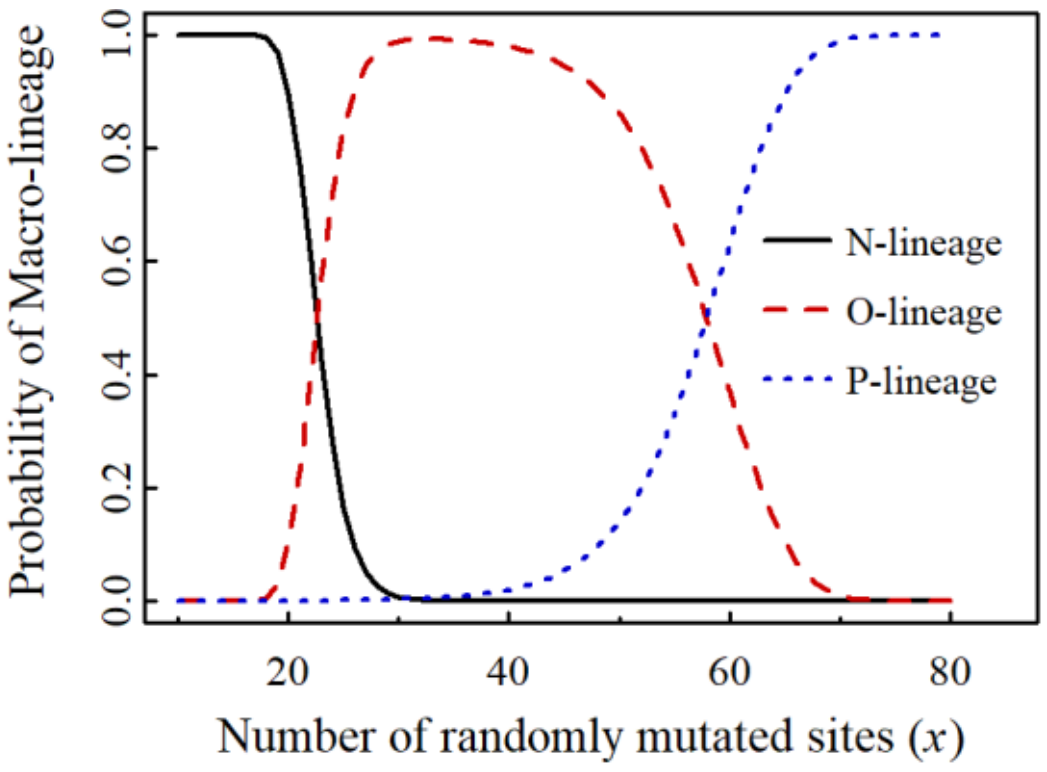

**Fig. 1 PML(Probability of Macro-lineage) vs $x$ for 25 mutants ($a$=104) given in Table 1**

The calculation uses a 99% percentile. We found that the assignment of a macro-lineage to a new variant depends statistically on $x$, with specific demarcation values distinguishing different macro-lineages as follows:

$New_{x.y} \in$ N when $x\leq 18$; $New_{x.y} \in$ N or O when $18<x\leq 29$; $New_{x.y} \in$ O when $29<x\leq 36$; $New_{x.y} \in$ O or P when $36<x\leq 69$; and $New_{x.y} \in$ P when $x>69$.

At a later time, $T_2$ (July 20, 2024), after incorporating 11 new mutants (listed in Table S2) into set $A$, the value of $a$ increased from 104 to 128, covering a total of 36 mutants. We again performed $10^5$ rounds of stochastic sampling for $x$ and 6 rounds of randomization for $y$. The probabilities of the $j$-lineage ($j$ = N, O, P, or Q) are shown in Fig. 2.

Using the same method, we obtained the following demarcation values for $x$ in the case of 36 mutants:

$New_{x.y}$ ∈ N when $x$≤18; $New_{x.y}$ ∈ N or O when 18<$x$≤ 29; $New_{x.y}$ ∈ O when 29<$x$≤ 39; $New_{x.y}$ ∈ O or P when 39<$x$≤62; $New_{x.y}$ ∈ P when 62<$x$≤72; $New_{x.y}$ ∈ P or Q when 72<$x$≤ 99 and $New_{x.y}$ ∈ Q when $x$>99.

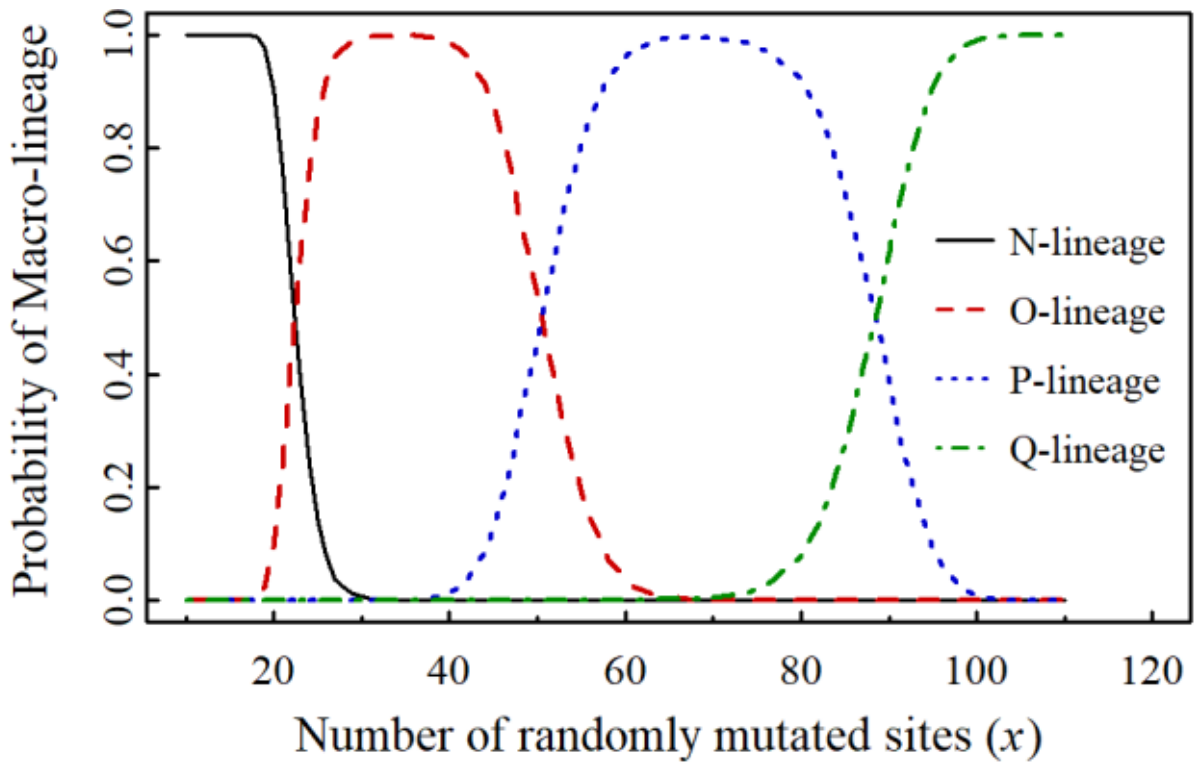

**Fig. 2 PML(Probability of Macro-lineage) vs *x* for 36 mutants (*a*=128) given in Table 2**

At the next time point, $T_3$ (January 4, 2025), after incorporating 70 variants from Table 3 into set $A$, the value of $a$ increased to 153. The probabilities of the $j$-lineage (where $j$ = N, O, P, or Q) for 70 mutants are presented in Fig. 3. The demarcation values for $x$ are:

$New_{x.y}$ ∈ N when $x$≤21; $New_{x.y}$ ∈ N or O when 21<$x$≤ 30; $New_{x.y}$ ∈ O when 30<$x$≤ 37; $New_{x.y}$ ∈ O or P when 37<$x$≤61; $New_{x.y}$ ∈ P when 61<$x$≤84; $New_{x.y}$ ∈ P or Q when 84<$x$≤ 112 and $New_{x.y}$ ∈ Q when $x$>112.

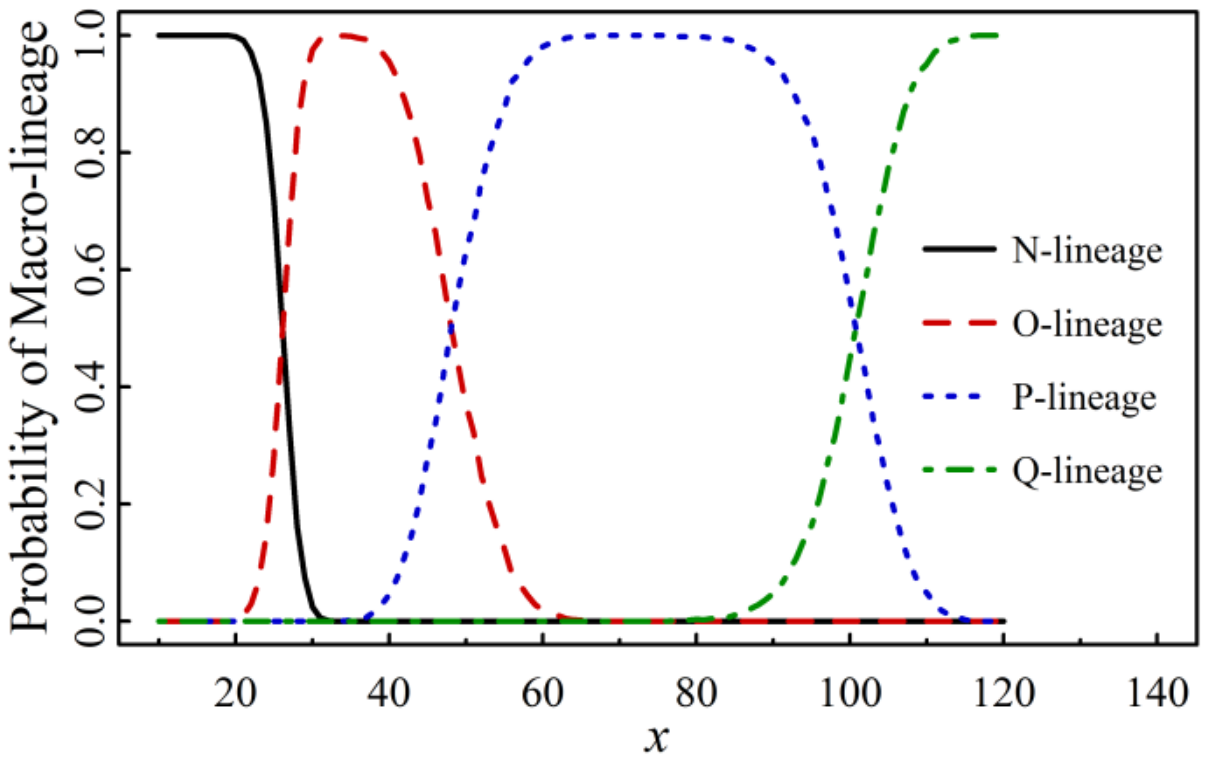

**Fig. 3 PML(Probability of Macro-lineage) vs *x* for 70 mutants (*a*=153) given in Table 3**

## *Law for the emergence of novel macro-lineages*

Set $A(a)$ is an observable set that depends on time $t$. Set $X$ represents the set of assumed mutated sites, also dependent on $t$, used to predict new macro-lineages. We define $x_{\text{dem1}}$ as the demarcation value of $x$ at which the occurrence of a new macro-lineage has a nonzero probability (greater than 1% in calculations), and $x_{\text{dem2}}$ as the demarcation value for its definitive occurrence. Both $x_{\text{dem1}}$ and $x_{\mathrm{dem2}}$ are functions of time $t$.

At time $t_k$, let the value of $a$ be denoted as $a_k$, with examples $a_1$ = 104, $a_2$ = 128 and $a_3$ = 153. Denote $x_{\mathrm{dem1}}$ for macro-lineage $j$ (where $j$ = O, P, Q) at time $t_k$ as $x^{(j)}_{\mathrm{dem1}}(t_k)$, and similarly, denote $x_{\mathrm{dem2}}$ for macro-lineage $j$ at time $t_k$ as $x^{(j)}_{\mathrm{dem2}}(t_k)$. The values of $x^{(j)}_{\mathrm{dem1}}(t_k)$ and $x^{(j)}_{\mathrm{dem2}}(t_k)$ are shown in Figs. 1, 2 and 3, and summarized in Table 4.

**Table 4 Demarcation values $x_{\mathrm{dem1}}$ and $x_{\mathrm{dem2}}$ at three times $t_1$ and $t_2$ and $t_3$**

| $k$ | $a_k$ | $x^{(O)}_{\mathrm{dem1}}(t_k)$ | $x^{(O)}_{\mathrm{dem2}}(t_k)$ | $x^{(P)}_{\mathrm{dem1}}(t_k)$ | $x^{(P)}_{\mathrm{dem2}}(t_k)$ | $x^{(Q)}_{\mathrm{dem1}}(t_k)$ | $x^{(Q)}_{\mathrm{dem2}}(t_k)$ |
|---|---|---|---|---|---|---|---|
| 1 | 104 | 19 | 30 | 37 | 70 | - | - |
| 2 | 128 | 19 | 30 | 40 | 63 | 73 | 100 |
| 3 | 153 | 22 | 31 | 38 | 62 | 85 | 113 |

Note: $t_1 < t_2 < t_3$. $t_1$ is the time period later than $T_1$=October 15, 2023, $t_2$ is the time period later than $T_2$= July 20, 2024, $t_3$ is the time period later than $T_3$= January 4, 2025.

Table 4 outlines the occurrence pattern for novel macro-lineages: the O-lineage emerges at $x \approx$ 19 - 22, and definitely occurs at $x \approx$ 30 - 31; the P-lineage emerges at $x \approx$ 37 - 40, and definitely occurs at $x \approx$ 62 - 70; the Q-lineage emerges at $x \approx$ 73 -85, and definitely occurs at $x \approx$ 100 - 113. The demarcation values deduced in different time periods are highly consistent.

# Part 3 Mutation Sites Increase over Time in SARS-CoV-2 Variants

**Table 5 Retrieval of SARS-CoV-2 variant data in Table 3: The number of mutated sites (NMS) in the spike protein and the first global sample collection dates for each variant arranged in chronological order**

| Macro-lineage | Variant | NMS * | Earliest date ‡ | Variant | NMS * | Earliest date ‡ |
|---|---|---|---|---|---|---|
| N-lineage | B.1 | 1 | 15 Jan 2020 | B.1.621 | 9 | 19 Sep 2020 |
| | B.1.177 | 2 | 7 Mar 2020 | R.1 | 4 | 24 Oct 2020 |
| | B.1.1.176 | 3 | 12 Mar 2020 | C.37 | 14 | 8 Nov 2020 |
| | D.2 | 2 | 19 Mar 2020 | B.1.526 | 4 | 15 Nov 2020 |
| | P.2 | 3 | 15 Apr 2020 | B.1.525 | 9 | 11 Dec 2020 |
| | B.1.1.7 | 10 | 14 May 2020 | C.36.3 | 9 | 4 Jan 2021 |
| | B.1.429 | 4 | 6 Jul 2020 | P.3 | 7 | 15 Jan 2021 |
| | B.1.351 | 10 | 9 Jul 2020 | AZ.2 | 6 | 5 Feb 2021 |
| | B.1.617.2 | 9 | 7 Sep 2020 | AV.1 | 10 | 23 Mar 2021 |
| | P.1 | 12 | 11 Sep 2020 | B.1.1.529 | 7 | 15 Apr 2021 |
| | B.1.617.1 | 5 | 15 Sep 2020 | C.1.2 | 15 | 11 May 2021 |
| O-lineage | BA.1 | 33 | 27 Jan 2021 | BN.1.2 | 40 | 7 Feb 2022 |
| | BA.1.1 | 35 | 28 Jan 2021 | CH.1.1 | 41 | 12 May 2022 |
| | BA.2 | 31 | 25 Mar 2021 | XBB.1.5 | 42 | 12 Jun 2022 |
| | BA.2.12.1 | 33 | 28 Sep 2021 | BM.4.1.1 | 39 | 20 Jul 2022 |
| | BA.2.65 | 31 | 11 Oct 2021 | XBC.1 | 39 | 6 Sep 2022 |
| | BA.1.1.15 | 37 | 27 Nov 2021 | CH.1.1.1 | 42 | 15 Oct 2022 |
| | BA.5 | 34 | 9 Dec 2021 | XBB.1.16 | 43 | 4 Jan 2023 |
| | BA.4.1 | 35 | 14 Dec 2021 | EG.1 | 43 | 16 Jan 2023 |
| | BQ.1.1 | 37 | 20 Dec 2021 | HV.1 | 46 | 29 Jan 2023 |
| | BA.2.75 | 30 | 31 Dec 2021 | HK.3 | 45 | 29 Jan 2023 |
| | BF.5 | 35 | 8 Jan 2022 | EG.5.1 | 44 | 31 Jan 2023 |
| | BF.7 | 35 | 24 Jan 2022 | DV.7.1 | 45 | 29 May 2023 |
| P-lineage | JN.1 | 60 | 13 Jan 2023 | KP.3 | 63 | 7 Jan 2024 |
| | BA.2.86.1 | 59 | 17 Jan 2023 | LB.1 | 64 | 15 Jan 2024 |
| | BA.2.86 | 58 | 11 Mar 2023 | KP.1 | 63 | 1 Feb 2024 |
| | JN.2 | 59 | 22 Jun 2023 | KS.1 | 58 | 15 Feb 2024 |
| | JN.1.7 | 62 | 25 Sep 2023 | KP.1.1.3 | 65 | 23 Feb 2024 |
| | JN.1.18 | 57 | 11 Dec 2023 | XDV.1 | 56 | 26 Feb 2024 |
| | JN.1.11.1 | 62 | 29 Dec 2023 | LP.1 | 66 | 22 Apr 2024 |
| | KP.3.1.1 | 64 | 1 Jan 2024 | XED | 64 | 19 Jun 2024 |
| | KP.2 | 59 | 2 Jan 2024 | XEC | 65 | 28 Jun 2024 |
| | JN.1.37 | 61 | 3 Jan 2024 | LF.7 | 67 | 26 Aug 2024 |
| | XEB | 61 | 3 Jan 2024 | LF.7.1.2 | 68 | 8 Sep 2024 |
| | XDQ.1 | 55 | 5 Jan 2024 | LP.8.1 | 68 | 19 Sep 2024 |

* NMS: number of mutated sites. ‡ Dates : the worldwide first sample collection date.

***Each macro-lineage of SARS-CoV-2 has a specific survival time. The relationship between the number of mutated sites and time t is a discontinuous function***

In Table 5 by the retrieval of SARS-CoV-2 variant data in Table 3, the relation between the number of mutated sites (NMS) in the spike protein and the first global sample collection dates for each variant is given and arranged in chronological order. Our findings show that NMS ($t$) is a discontinuous function of time, corresponding to three distinct macro-lineages, as illustrated in Fig. 4a. Fig. 4a is based on the data presented in Table 5.

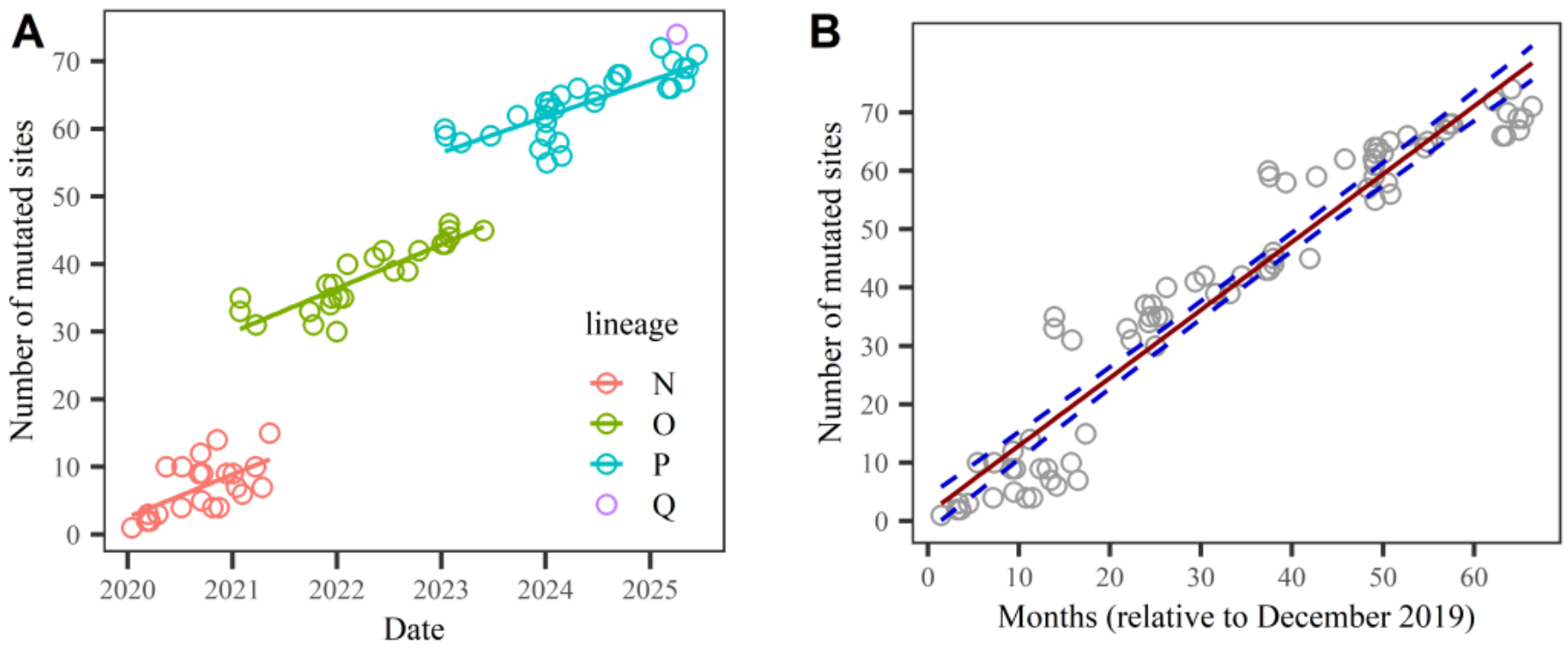

**Fig. 4** (A)Number of mutated sites as a function of the first sample collection date. (B) Linear regression of the number of mutated sites versus the first sample collection date. The 95% confidence intervals were calculated using linear regression theory and represented by the blue dotted lines

Furthermore, we performed linear regression to examine the relationship between the number of mutated sites and the first sample collection date, with the results presented in Fig.4B. The linear regression model effectively approximates the increasing trend in the number of mutated sites over the course of viral evolution. The $R^2$ and RSE for the linear regression of NMS are 0.92 and 6.68, respectively. Additionally, applying linear regression theory, we determined the slope of the regression line equaling 1.163 per month, with a 95% confidence interval of ±0.076. This linear regression provides a better fit, and we will use the linear relationship between the number of mutated sites ($x$) and collection time ($t$) to predict the emergence of new viral lineages.

By extending the linear relationship over a longer period and using linear regression theory, we forecast that the number of mutated sites will reach NMS =73.4 ±2.6 to 85 ± 3 by the 62th to 72th month and 100 ± 4 to 113 ± 5 by the 85$^{th}$ to 96th month, starting from December 2019. Combining this with the demarcation values $x_{dem1}$ = 73-85 (indicating the initial emergence of Q) and $x_{dem2}$ = 100-113 (indicating a strong outbreak of Q), we predict that the macro-lineage Q will emerge around February to December 2025, and after approximately 22 to 24 months, may experience a strong outbreak. However, due to the mutation probabilities of various sites on spike protein largely influenced by all kinds of biological driving forces the real timeline of new macro-lineage emergence should be altered and the above preliminary estimation is for reference only. At present, it appears that the new variant BA.3.2, called Cicada, was generated in 2025 June and now spreading in USA . Perhaps it is a signal for the emergence of the new macro-lineage Q.